\documentstyle[aps,epsfig,multicol]{revtex} 
\draft 
 
\begin{document} 
\author{Lin Tang} 
\address{Department of Physics, Nanjing University, Nanjing 210093, China} 
\title{Effect of histamine on the electric activities of cerebellar 
Purkinje cell} 
\date{January, 1999} 
\maketitle 
 
\begin{abstract} 
The effect of histamine (HA) on the electric activities of Purkinje cell 
(PC) is studied on the cerebellum slice. We find that: (1) HA's main effect 
on PC is excitative (72.9\%); there are also a small amount of PC showing 
inhibitive (10.2\%) or no (16.9\%) response to HA. (2) Different from the 
conventional opinion, HA's excitative effect on PC is mutually conducted by 
H1 and H2 receptors; the antagonist for H1 receptor could weaken HA's 
excitative effect on PC, while the antagonist for H2 receptor could weaken 
or even block the excitative effect of HA on PC. (3) PC's reaction to HA is 
related to its intrinsic discharge frequency; there exists a frequency at 
which PC is highly sensitive to HA, and well above this frequency PC becomes 
stable against HA. These results indicate that the histaminergic afferent 
fibre can adjust PC's electric activities by releasing HA, and thereby 
influence the global function of the cerebellar cortex; and that just like 
the $\gamma $ region of cerebrum, cerebellum may also have some sort of 
characteristic frequency. 
 
\pacs{PACS number: 87.10.+e} 
 
\noindent{Key words: cerebellar cortex; Purkinje cell; histamine; receptor; 
                     cerebellum slice} 
 
\end{abstract} 
 
\begin{multicols}{2} 
 
\section{Introduction} 
 
Cerebellum deserves more extensive studies than was conventionally realized. 
Historically, cerebellum was thought of as mainly a motor control organ, 
while recent researches reveal that it has many other functions, and it has 
more intimate connections to other parts of the brain \cite{non-motor}. 
Histamine (HA), a neurotransmitter or 
neuromodulator in the brain, plays an important role in the functions and 
interactions of various parts of the brain, and also in the studies of these 
functions and interactions. For instance, the neuroanatamic researches 
revealed the existence of the hypothalamus-cerebellum histaminergic path  
\cite{Schwartz,Li}, and shows that hypothalamus has great influence on the 
cerebellar activities and hence plays an important role in coordinating the 
functions of the body and viscera. Besides, HA has the possible role of 
controlling the cerebellar circulation, and HA receptors are also found in 
the neurons of the cerebellar cortex \cite{Schwartz,Li}. 
 
Cerebellum is made up of the outer dark matter (cortex), inner white matter, 
and three pairs of deep nuclei lying in the heart of the white matter. These 
nuclei are the fastigial nucleus FN, interposed nucleus IN, and dentate 
nucleus DN. The afferent fibres to the cerebellum are mainly from the 
vestibule, spinal cord, and the cerebral cortex. They form synaptic 
connections with the neurons in the cerebellar deep nuclei and cortex. The 
synapses of Purkinje cell (PC) make the efferent fibres of the cerebellar 
cortex, they are mainly projected into the deep nuclei, then the neurons 
there stick out fibres which make the cerebellar output; a small amount of 
PC synapses are directly projected into the vestibular nucleus. 
 
The cerebellar cortex can the divided into three layers: the (out-most) 
molecular layer, PC layer, and the granular layer. It contains three kinds 
of afferent fibres (musciform fibre MF, crawl fibre CF, and monoaminergic 
fibre), and five kinds of neurons (PC, granular cell GR, basket cell BA, 
star-like cell ST, and Golgi cell GO). Hence we see that the cerebellar 
afferent fibres and the intermediate neurons, with PC acting as the core, 
constitute the basic neural circular that is responsible for the sensory 
function of the cerebellar cortex, and the cerebellar cortex together with 
the deep nuclei undertake various functions of the cerebellum. 
 
In our laboratory there has been studies of HA's effects on certain neurons 
of the cerebellum, such as the granular cells \cite{Li,Li'}, while HA's 
effect on the neurons in the cerebellar cortex is not yet intensively 
studied. Considering that PC is the only efferent neuron of the cerebellar 
cortex, we are going to investigate HA's effect on the electric activities 
of PC, based on our accumulated experiences in studying the influences of 
aminergic materials such as norepinephrine (NA) and serotonin (5-HT) on the 
spontaneous and induced discharge activities of the cerebellar PC; so as to 
learn more about the role of aminergic afferent system in the process of 
information treatment in the cerebellar cortex. 
 
\section{Experimental material and method} 
 
We use for our experiments 19 mature SD rats (200-250g). Anaesthetize a rat 
by injecting betchloramines hydrochloride (4mg/100g) into the abdominal 
cavity, take out the cerebellum right after cutting the head, wash the 
cerebellum with frozen artificial cerebrospinal fluid (ACSF, $4^{\circ }$C), 
stick it onto the operating table of a microtome (at which the cerebellum is 
soaked in frozen ACSF), and cut a 400${\rm \mu m}$ thick of arrow-like slice 
from the vermis. The process of making the slice should be done within 20 
minutes. Then move the so obtained slice into a recording trough, and begin 
the experiment after 15 minutes of hatching. The recording trough is 
continuously irrigated (3ml/min) by ACSF (33$\pm 0.2^{\circ }{\rm C}$), and 
is aerated with the mixed air of 95\%O$_2$ +5\%CO$_2$. The concentrations 
of various elements in ACSF are (mmol/l): NaCl 124, KCl 5, KH$_2$PO$_4$ 1.2, 
MgSO$_4$ 1.3, CaCl$_2$ 2.4, NaHCO$_3$ 26, glucose 10. Put a tiny glass 
electrode (filled in with colored conducting fluid) at the PC 
layer of the X leaflet 
of the cerebellar cortex to make out-cell records of PC's discharge 
activity. This is because the X leaflet received the least mechanical wound 
in making the slice, and hence the cerebellum slice should be so put in the 
recording trough that the X leaflet be well hatched by the ACSF and the 
mixed air. 
 
We base on the following criteria to single out PC discharge signals: (1)  
{\it position of the electrode}: PCs in the cerebellar cortex are of linear 
type, namely, the cell bodies concentrate at one end and the dendrites stick 
to the other end, thereby form a PC layer in the cerebellar cortex. We put 
the recording electrode at the out side of the PC layer and near to the 
molecular layer, therefore there is very little chance to catch a discharge 
signal of the granular cell; (2) {\it discharge wave shape}: the PC action 
potential has large magnitude (0.700$\pm $0.143${\rm \mu V}$) 
and thick contour \cite 
{Huang}, moreover, the effective recording distance of PC discharge is 50$%
{\rm \mu }$m, much longer than that of the granular cell (20-30${\rm \mu }$%
m) \cite{Li'}; (3) {\it cell numbers}: in the cerebellar cortex there are 
much less GO, ST, and BA than PC, and hence the discharge signals from these 
cells may be ignored. 
 
Because the nerve cells in the brain have the character of continuously 
producing impulses, and often at the frequency of tens of times per second, 
the physiological status of a neuron is usually symbolized by its discharge 
frequency. Under the environmental stimulation, the discharge frequencies of 
central neurons will variate, and the intrinsic potentials will also 
deviate. Hence we magnify the PC discharge signal, choose a relatively small 
time constant for the recording system, so as to differentiate the input 
signal and transform the slowly increasing wave into a sharp high peak wave, 
then use a pulse discriminator to convert the peak signal into TTL pulse and 
input it into the computer, and then draw the post-stimulus histogram. Thus 
we can describe PC's reaction to chemicals by the post-stimulus histogram of 
its discharge frequency. 
 
All the drugs here are contemporarily compounded using ACSF, and are used to 
irrigate the cerebellum slice separately. When doing the experiments of 
studying the receptor mechanism of HA's effect on PC, the slice is 
continuously irrigated by the receptor antagonist for over 10 minutes before 
by HA. 
 
\section{Results} 
 
\subsection{The effect of HA on PC's electric activities} 
 
Because intermittent discharge is often due to the ill state of PC or 
indistinguishability of cell discharge signal from other signals, we choose 
in 29 cerebellum slices 59 cells with relatively stable spontaneous 
discharge for our study. Their spontaneous discharge frequencies range from 
4.48 to 78.32 Hz. We find that of the 59 cells, 83.1\% (49/59) are affected 
by HA, the other 16.9\% (10/59) show no response. And of the 49 responsive 
ones, 87.8\% (43/49) are excited, the other 12.2\% (6/49) are inhibited. 
(See Fig. 1) Therefore HA's effect on PC is mainly excitative. From Figs. 
1B, 1C, we can see that there is an evident increase of the variation of 
the cell's discharge frequency as the irrigating concentration of HA 
increases. 

\begin{minipage}{8.5cm}
\begin{figure}[tb]

\begin{tabular}{l}
\hspace{1cm} \psfig{figure=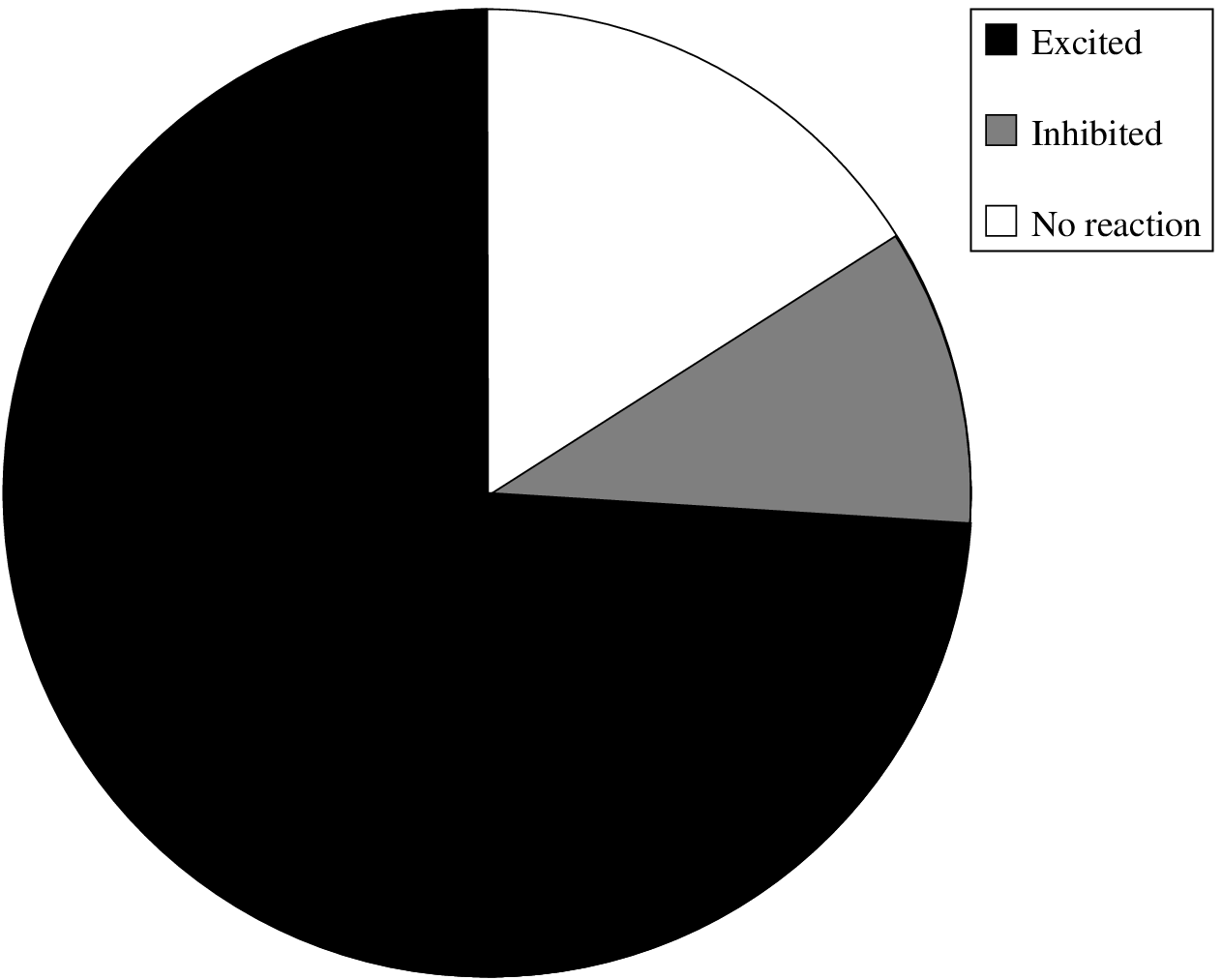,width=50mm}
\end{tabular}
\hspace{1.3cm}{\Large ~A}

\begin{tabular}{l}
\psfig{figure=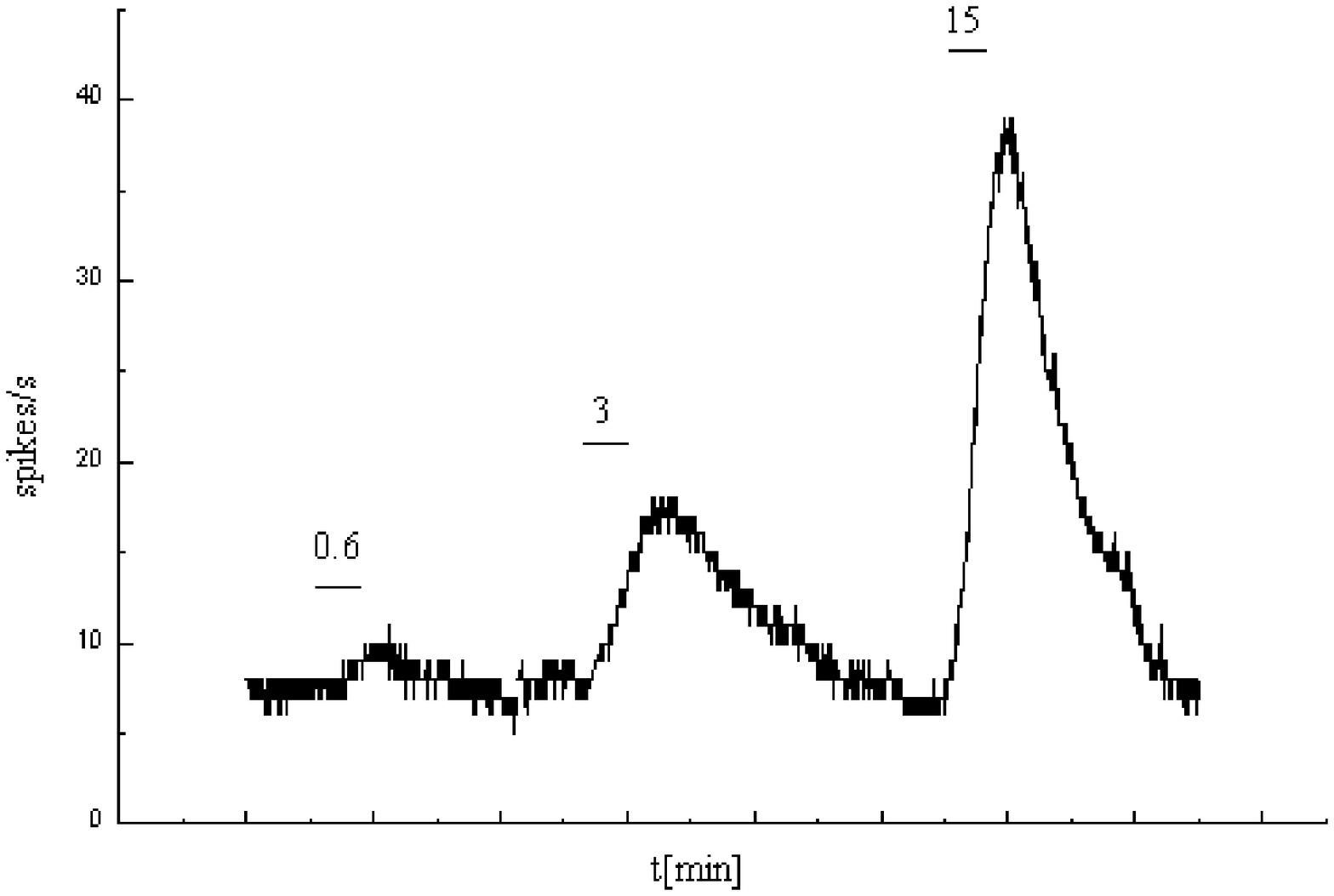,width=7.5cm}
\end{tabular}
{\Large ~B}

\begin{tabular}{l}
\psfig{figure=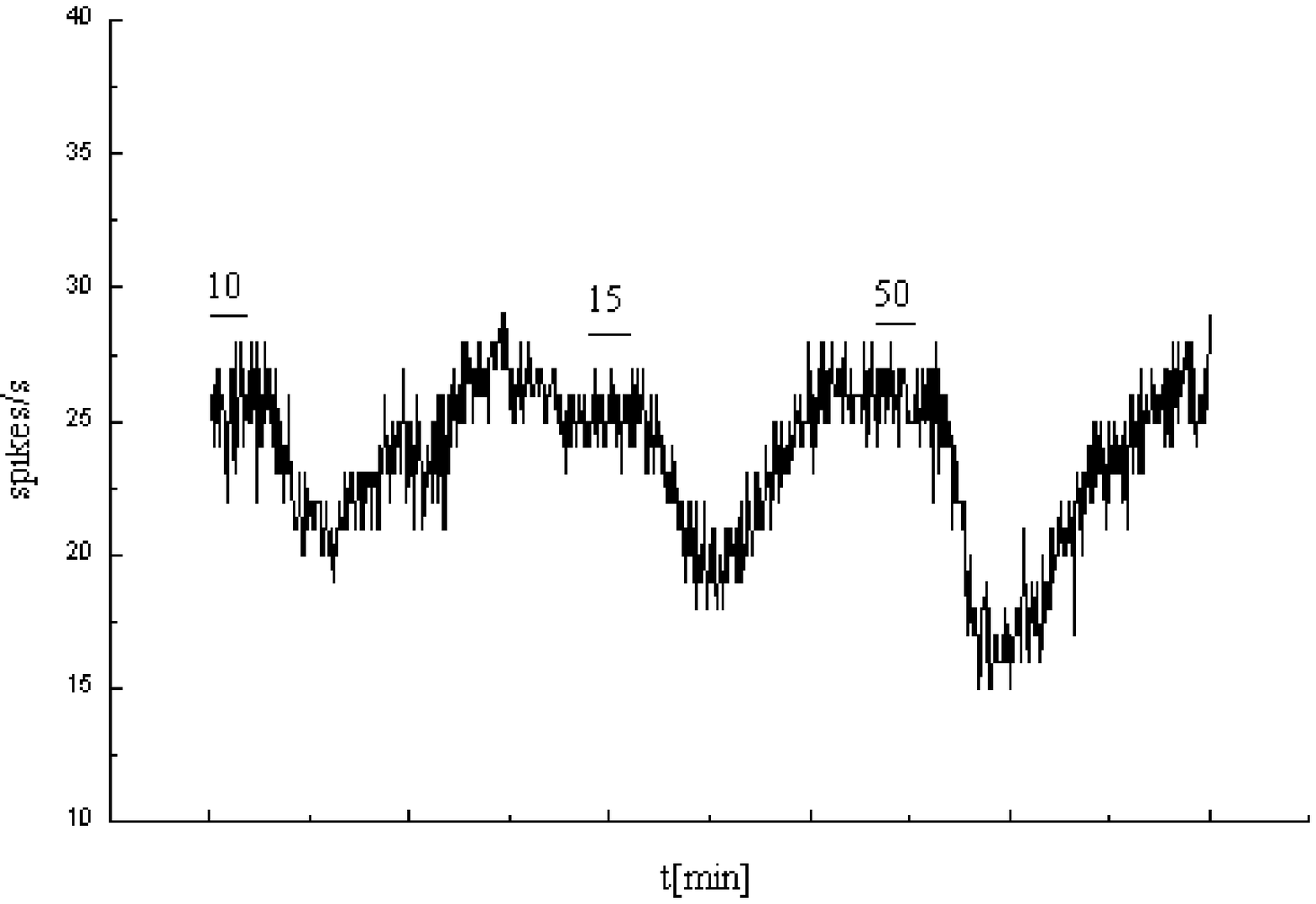,width=7.5cm}
\end{tabular}
{\Large ~C}

\caption{HA's effect on PC; 
(A) shows the percentage of PC's various reactions to HA; 
(B) and (C) show HA's excitative and inhibitive effects on PC, respectively; 
numbers above the bar represent HA's irrigating concentration.} 

\end{figure}
\end{minipage}

\bigskip 

Another important observation is that whether or not a PC shows response to 
HA may have some connection with its spontaneous discharge frequency. (cf. 
Fig. 2) The spontaneous discharge frequencies of the 59 tested cells are 
4.48-78.32 Hz, while that of the 49 cells which show response to HA is 21.34$%
\pm $12.69 Hz (M$\pm $SD), this is significantly lower than the average 
frequency (37.00$\pm $7.73 Hz) of the 10 cells unaffected by HA. However, 
whether a cell is excited or inhibited by HA seems to have nothing to do 
with its intrinsic frequency. The average frequencies of the 43 excited 
cells and of the 6 inhibited cells are 21.53$\pm $13.20 and 20.00$\pm $8.80 
Hz, respectively, which are not much different (P$>$0.5, 
t test). This differs from how HA's effect on the cerebellar granular 
cells is related to their intrinsic frequencies \cite{Li}. We also find that 
among the 49 responsive cells, those who have higher intrinsic discharge 
frequencies are more sensitive to HA (there are 8 responsive cells with 
intrinsic frequencies higher than 30 Hz, and 7 of them show response to HA 
at the lowest HA concentration of less than 30${\rm \mu }$mol/l). 

\begin{minipage}{8.5cm}
\begin{figure}[tb]

\begin{tabular}{l}
\psfig{figure=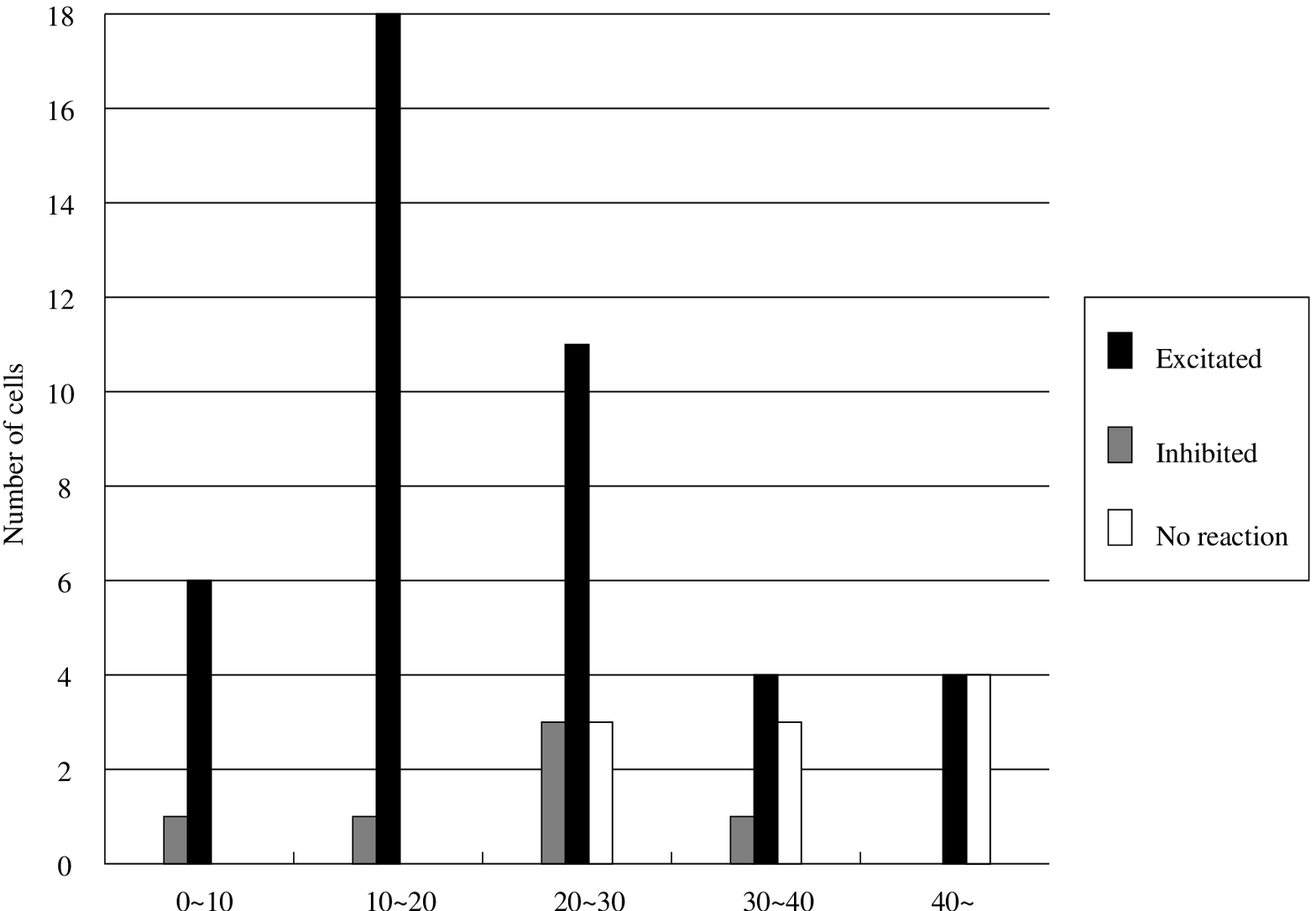,width=7.5cm}
\end{tabular}
{\Large ~A}

\vspace{5mm}

\begin{tabular}{l}
\psfig{figure=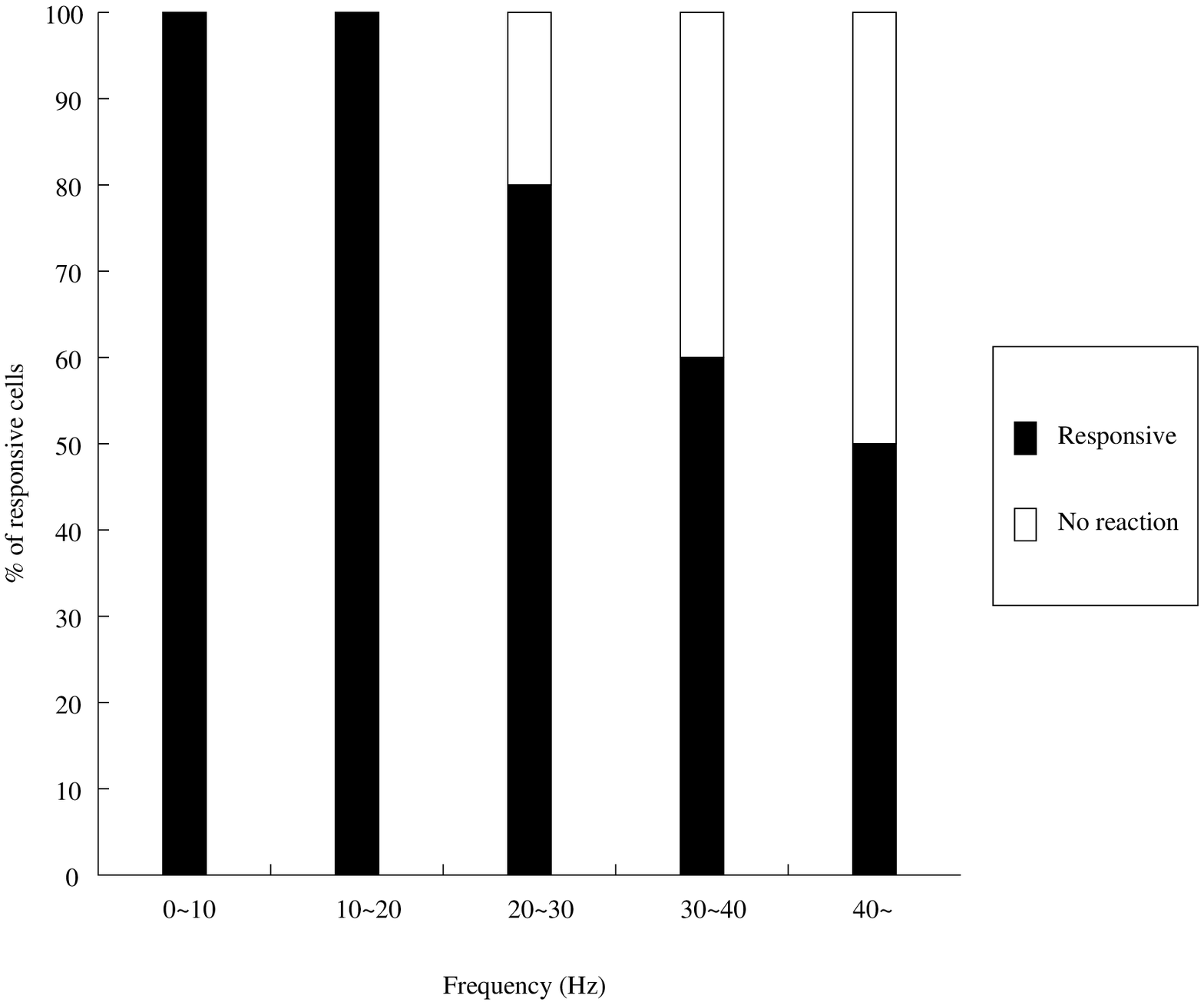,width=7.5cm}
\end{tabular}
{\Large ~B}

\caption{PC's discharge frequency and reaction to HA}

\end{figure}
\end{minipage}
 
\subsection{The receptor conducting mechanism of HA's excitative effect on PC 
} 
 
It is known that there are three sub-types of HA receptors in brain: H1, H2, 
and H3, therefore it is necessary to study the receptor conducting mechanism 
of HA's effect on PC. Since HA's main effect on PC is excitative, we make 
study for the excitative effect. For this purpose we observe the influences 
of Triprolidine (hyper-specific antagonist for H1 receptor) and Ranitidine 
(hyper-specific antagonist for H2 receptor) on PC's excitative reaction to 
HA, and find that low-concentration Triprolidine (0.5-1.0${\rm \mu }$mol/l) 
could weaken HA's excitative effect on PC, while low-concentration 
Ranitidine (0.9-1.0${\rm \mu }$mol/l) could weaken or even block HA's 
excitative effect on PC. (cf. Fig. 3). 
 
H3 receptor is a presynaptic self-receptor, it adjusts the presynaptic 
release of HA \cite{Schwartz}. In this research we have not investigated its 
role in HA's excitative effect on PC.

\begin{minipage}{8.5cm}
\begin{figure}[tb]

\begin{tabular}{l}
\psfig{figure=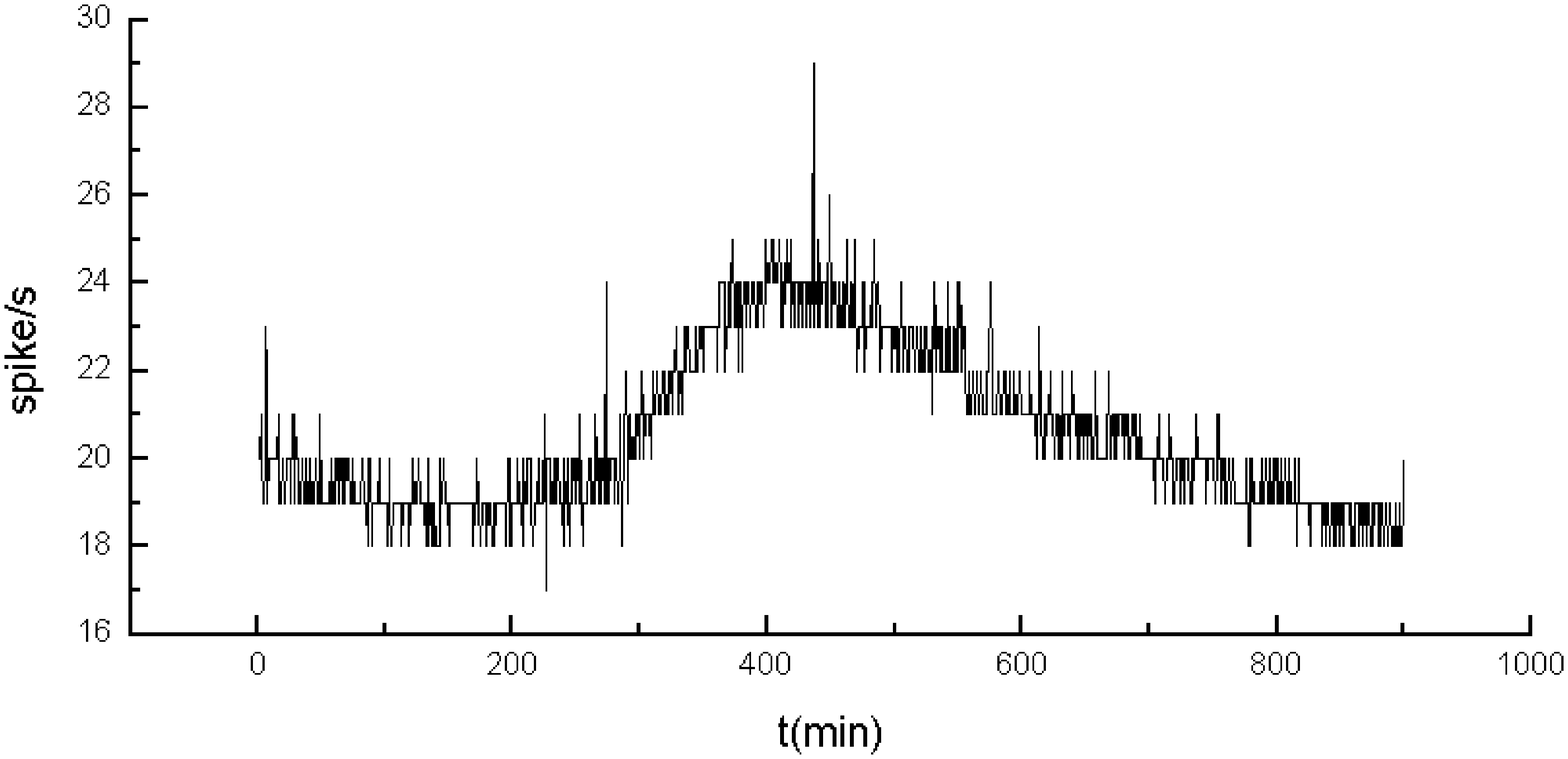,width=7.5cm}
\end{tabular}
{\Large ~A}

\begin{tabular}{l}
\psfig{figure=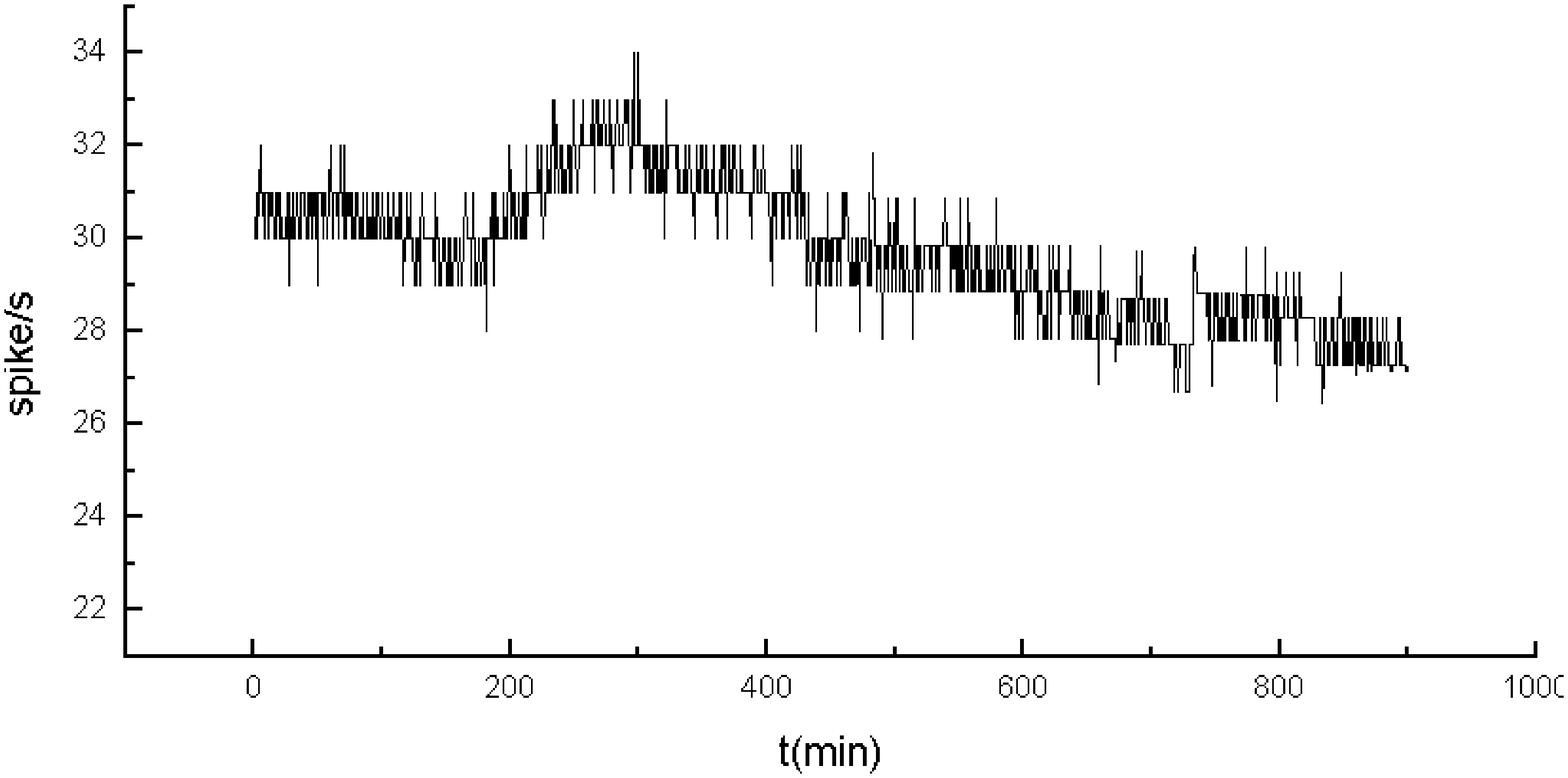,width=7.5cm}
\end{tabular}
{\Large ~B}

\begin{tabular}{l}
\psfig{figure=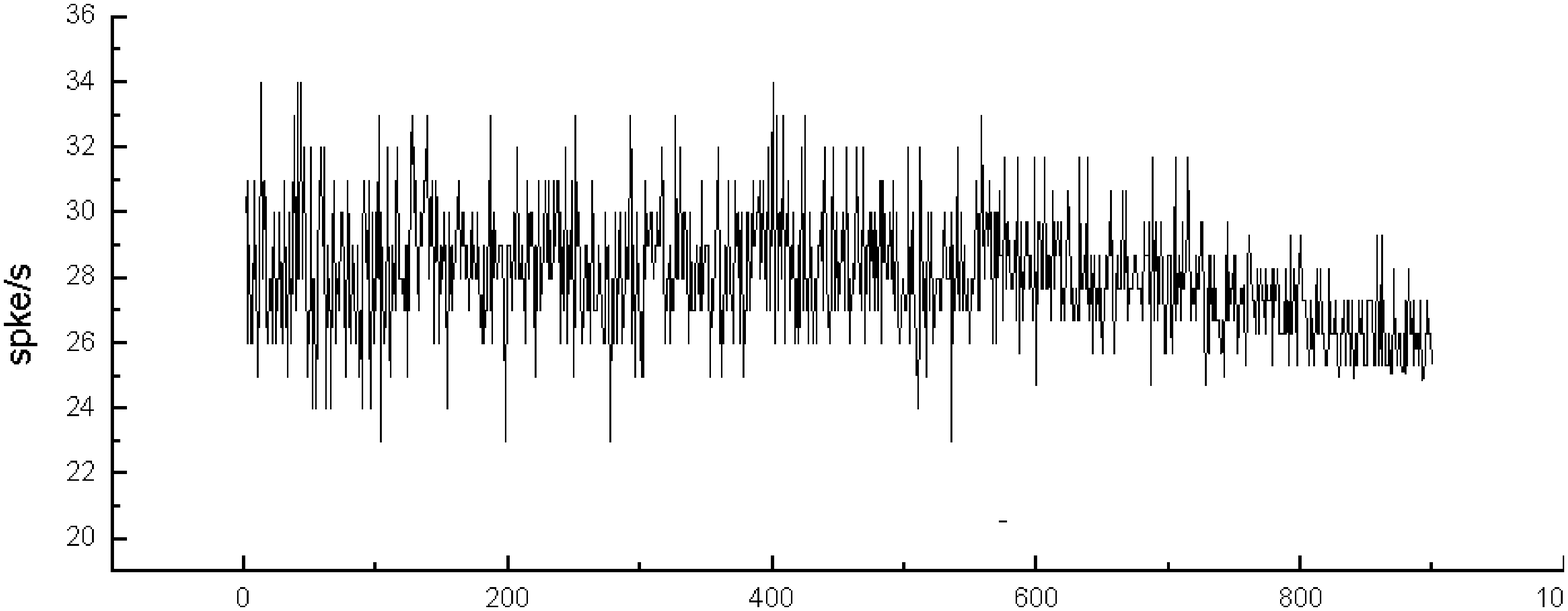,width=7.5cm}
\end{tabular}
{\Large ~C}

\caption{Influence of receptor antagonist on HA's excitative effect on PC;
(A) shows the normal excitative effect of HA (50${\rm \mu}$mol/l) on PC; 
(B) shows that after irrigating PC with ACSF containing 0.5${\rm \mu}$mol/l of
 Triprolidine for 15 minutes, HA's excitative effect on PC is significantly 
weakened; 
(C) shows that after irrigating PC with ACSF containing 0.9${\rm \mu}$mol/l of
Ranitidine, HA's excitative effect on PC is blocked.}    

\end{figure}
\end{minipage}
 
\section{Discussions} 
 
\subsection{Data-taking system} 
 
In this experiment we used the data-taking system: oscillometer-pulse 
discriminator-computer, and drawn the post-stimulus histogram, then evaluate 
PC's reaction to chemicals by the post-stimulus histogram of the discharge 
frequency. The sampling bin width is 3, point=900, interval=1000, one data 
represents the discharge times of a cell in one second. Since a cell 
discharges at very high frequency and its electric signals transmit at very 
high speed, it gives very rich information in one second. We know that the 
information carried by a neuron is represented by its frequency distribution 
over the action potential, therefore if we can obtain the frequency 
spectrum, and make the combined analyses of the frequency and power spectra, 
we will get greater amount information. Besides, what we take in this 
experiment is out-cell record of a cell's discharge activities, and PC is in 
connection with various other cells in the cerebellum slice, the measurement 
in such case may be disturbed by the environment, and the excitative or 
inhibitive reaction of PC may actually be a combined result of various 
influences including those from the environment. These are to be improved in 
future studies. 
 
\subsection{HA's effect on PC} 
 
Histamine is an important neurotransmitter or neuromodulator, and acts as an 
inter-cell messenger. HA can hardly penetrate the blood-brain barrier, hence 
must be produced by the histaminergic neurons. It can be released through 
depolarization and by means of calcium dependence. The 
hypothalamus-cerebellum histaminergic fibres in the brain are casted by the 
hypothalamus tuberous mastoid nucleus (HA pericaryon) into the cerebellar 
cortex and the cerebellar deep nuclei (HA nerve end) \cite{Schwartz,Li}. 
Cerebellum-hypothalamus histaminergic system can bring about many functions 
via the inter-cell messenger HA. The hypothalamus-cerebellum histaminergic 
projecting fibres end at the cerebellar cortex as spreading multi-layer 
fibres, and their nerve ends mainly spreadingly adjust the functions of the 
peripheral neurons in the varix form; these are much like the serotoninergic 
and norepinephrinergic afferent fibres of the cerebellum \cite{Schwartz,Li'}. 
 
Purkinje cell is the only kind of efferent neuron of the cerebellar cortex, 
it is a deformation of the multi-pole cell, and has very strong dentrite. Our 
research shows that PC's discharge activity is significantly affected by 
HA. From our observations in Sec. IIIA, and associating how the reaction of 
cerebellar granular cells to HA is related to their intrinsic discharge 
frequencies \cite{Li}, 
we speculate that the cerebellar afferent fibres might be 
mainly responsible for adjusting the basic discharge levels of the 
cerebellar neurons: as the discharge frequency of a target cell lies in a 
certain region it loses sensitivity to HA; and as the discharge frequency 
lies out of this region the cell reacts to HA, and thereby its frequency 
tends to that region. Moreover, the high-frequency region is a 
low-sensitivity region to HA, while for the responsive cells, those with 
higher frequency are highly sensitive to HA. We therefore speculate that 
there exists a critical frequency, a cell with intrinsic frequency near to 
this critical point will significantly change its states by only a small 
amount of HA, while as its frequency goes well above this point, it becomes 
stable. In connection with the influence of NA and 5-HA on the spontaneous 
and induced electric activities of cerebellar PC \cite{Wang}, we go further 
to speculate that histaminergic and other aminergic afferent systems might, 
through their coordinations and/or antagonism, adjust the synaptic 
transmission efficiency of MF-PF (parallel fibre)-PC and CF-PC, or adjust 
PC's sensitivity to signals from MF and CF, and thereby take part in the the 
global process of the sense movement of the cerebellar neuron net. That many 
kinds of neural active materials mutually act on the neuron is a common 
pattern of signal transmission of the nerve system; such mutual action may 
happen at the presynapse, postsynapse, or even postreceptor level. HA's 
effect on the cerebellar PC is to adjust the excitativity level of the 
neuron and the neuron's sensitivity to input information from outside of the 
cerebellum, and adjust the sleep or awake state of the cortex. The aminergic 
afferent fibres have synaptic or non-synaptic chemical transmission effect 
on the cerebellar cortex, their non-synaptic transmission may not encode the 
fast phase-like information, but hyperfinely adjusts the membrane potential 
and basic discharge level or the target neuron. 
 
\subsection{The receptor conducting mechanism of HA's excitative effect on PC 
} 
 
Of the three sub-types of HA receptors in the brain (H1, H2, and H3), H3 
receptor is a presynaptic self-receptor. The acting mechanism of H1 receptor 
is through some sub-type of G protein to activate PLC, hydrolize 4,5-PIP$_2$ 
to be 1,4,5-IP3 and DG, these two secondary messengers go further to trigger 
various biological effects, such as stimulating endoplasmic reticulum to 
release Ca$^{+}$and thence trigger the ion channels \cite{Schwartz}. The H2 
receptor brings effect mainly via G protein-AC-cAMP: HA combines with the H2 
receptor of the cell membrane, the H2 receptor is then deformed and triggers 
G protein to give off its $\beta $ and $\gamma $ sub-radicals, and on the $%
\alpha $ sub-radical GDP is replaced by GTP, then AC is activated and 
catalyzes ATP to transform to cAMP, cAMP acts as a secondary messenger and 
triggers a series of reactions, such as activating PKA and then triggering 
ion channels, or directly triggering ion channels. Because many highly 
efficient enzymes take part in these reactions, small signals are magnified 
step by step. 
 
It is usually thought that H1 and H2 receptors conduct HA's excitative and 
inhibitive effects on neurons respectively \cite{Schwartz}. However, this is 
in contradiction with our report in Sec. IIIB that low-concentration H1 
receptor antagonist could weaken HA's excitative effect on PC, while 
low-concentration H2 receptor antagonist could weaken or even block HA's 
excitative effect on PC. This makes us to conjecture that 
{\em both} H1 and H2 
receptors involve in HA's excitative effect on PC, and H2 receptor is the  
{\em main} conductor. There has been report that HA's excitative effect on 
the granular cells of the cerebellar cortex and on the cells of the vestibular 
inner-side nucleus is mutually conducted by H1 and H2 receptors \cite 
{Schwartz}. Histochemistry researches also reveal that in the cerebellar 
cortex of rat the density of H2 receptor is higher than that of H1 receptor. 
All these partly support our conjecture. This may suggest that the 
H2-related signal transmission chain in cell differs from one type of 
neuron to another, and such difference determines whether the H2 receptor 
conducts excitative or inhibitive effect; but the particular mechanism of 
signal transmission is still to be studied. 
 
\subsection{Mathematical simulation} 
 
To systematically study the biophysical or biochemical mechanisms of the 
reactions of cells, it is often helpful to make mathematical simulations of 
the reactions. Here we give a simulation for the relation between PC's 
(excitative) reaction to HA (measured by its maximum variation of discharge 
frequency under the action of HA at a certain irrigating concentration) and 
its intrinsic discharge frequency $f_0$. To be concrete, we choose a fixed 
frequency variation $A$, and denote by $u_0$ the irrigating concentration of 
HA at which the maximum change of a cell's frequency would be $A$, then 
study the relation between $u_0$ and $f_0$. For this purpose we first look 
at how the maximum frequency $f$ of a cell under the action of HA is related 
to HA's irrigating concentration $u$. 
 
As we noted in Sec. IIIA, PC's reaction to HA has something to do with its 
intrinsic discharge frequency: high-frequency cells are insensitive to HA, 
while for the cells affected by HA, the higher-frequency ones are highly 
sensitive to HA, and at low intrinsic frequencies $f$ increases steadily with  
$u$. We therefore assume that $f$ is an $S$-like function of $u$:  
\begin{equation} 
f(u)=f_c/(1+\exp (-(u-u_c))).  \label{f} 
\end{equation} 
As $u=0$, $f(0)$ is just the intrinsic frequency $f_0$:  
\begin{equation} 
f_0=f(0)=f_c/(1+\exp (u_c))\text{.}  \label{f0} 
\end{equation} 
According to our above explanations, we write  
\begin{equation} 
f(u_0)=f_0+A=f_c/(1+\exp (-(u_0-u_c))).  \label{fu0} 
\end{equation} 
From Eqs. (\ref{f0},\ref{fu0}) we can derive the $u_0$-$f_0$ relation:  
\begin{equation} 
u_0=\ln (f_c/f_0-1)-\ln (f_c/(f_0+A)-1)\text{.}  \label{u0} 
\end{equation} 
Here the physiological meaning of $f_c$ is the maximum discharge frequency 
of the cell, and $A$ is an arbitrarily chosen minimum frequency variation 
that can be observed in the experiment. It is straightforward to check that $%
u_0$ has a minimum value at an intermediate $f_0$, just as the experimental 
results show. Eq. (\ref{u0}) is much like the relation in the Hopfield 
model. But further study of the connection to the Hopfield model would 
require more detailed knowledge about the relation between $f$ and $u$, 
which is not possible in the present experiment, for PC cannot live through 
a long enough time to allow tests for various HA concentrations; hence the 
simulation here is merely a rough one. 
 
\subsection{Characteristic frequency in the cerebellum?} 
 
Eqs. (\ref{f},\ref{u0}) show a pattern very similar 
the characteristic frequency 
of the cerebrum (the $\gamma $ region of around 40 Hz): there exist a 
particular frequency at which the target cell is highly sensitive to HA, and 
well above this frequency the cell loses sensitivity to HA. For PC this 
insensitive region is 25.00-46.24 Hz, with the average 37.00$\pm $7.72, 
which is close to the $\gamma $ region. These suggest that cerebellum might 
also exhibit some sort of characteristic frequency to external stimulations. 
Further studies of such characteristic frequency would necessarily 
take into account of the 
interactions of various cell, such as in the HR model \cite{HR}; these we
hope to accomplish in the future.

\section{Acknowledgements} 
 
The author owes special thanks to Prof. Jiang-Jun Wang and Prof. Bing Zhu 
for valuable instructions, to Le Tian and Jie Ma for collaborations in this 
experiment, and to Feng Dong, Qin Xi, and Guo-Ning Hu for kind help and 
suggestions.

\end{multicols} 
 
\end{document}